\documentclass[preprint,showpacs,preprintnumbers,amsmath,amssymb]{revtex4}

\usepackage{graphicx}
\usepackage{dcolumn}
\usepackage{bm}

\begin{document}


\title{Bound states in a 2D short range potential induced by spin-orbit interaction}

\author{A.V.Chaplik
and  L.I.Magarill}
  \email{levm@isp.nsc.ru}
\affiliation{Institute of Semiconductor Physics, Siberian Branch
of the Russian Academy of Sciences, Novosibirsk, 630090, Russia}


\date{\today}

\begin{abstract}
We have discovered an unexpected and surprising fact: a 2D axially
symmetric short-range potential contains {\it infinite} number of
the levels of negative energy {\it if one takes into account the
spin-orbit (SO) interaction.} For a shallow well
($m_eU_0R^2/\hbar^2 \ll 1$, where $m_e$ is the effective mass,
$U_0$ and $R$ are the depth and the radius of the well,
correspondingly) and weak SO coupling ($|\alpha|m_eR/\hbar \ll 1$,
$\alpha$ is the SO coupling constant) exactly one two-fold
degenerate bound state exists for each value of the half-integer
moment $j=m+1/2$, and the corresponding binding energy $E_m$
extremely rapidly decreases with increasing $m$.
\end{abstract}

\maketitle

As it is well known from any textbook on quantum mechanics a very
shallow potential well ($m_eU_0R^2/\hbar^2 \ll 1$) cannot capture
a particle with the mass $m_e$ in 3D case and does this in 2D and
1D situations provided the wells are symmetric: even potential in
1D, axially symmetric well in 2D. In the latter case the only
negative level corresponds to the $s$-state ($m=0$).

Consider now a 2D electron with accounting for the SO interaction
in Bychkov-Rashba form \cite{bych}; the Hamiltonian reads:

\begin{equation}\label{Ham}
\hat{H} = \frac{\hat{p}^2}{2m_e} +
\alpha(\mbox{\boldmath{$\sigma$}}[\hat{\bf p}\times{\bf n}])+U(r),
\end{equation}
where $r$ and $\hat{\bf p}$ are the radius in cylindrical
coordinates  and  the 2D electron momentum operator, respectively,
$\mbox{\boldmath${\sigma}$}$ are Pauli matrices, ${\bf n}$ is the
normal to the plane of 2D system.

It is convenient to write down the Schr\"{o}dinger equation in the
${\bf p}$-representation:
\begin{equation}\label{Schreq_p}
[\frac{p^2}{2m_e} + \alpha(\mbox{\boldmath{$\sigma$}}[{\bf
p}\times{\bf  n}])]\Psi({\bf p})+\int \frac{d{\bf
p'}}{4\pi^2}{\cal U}({\bf p-p'})\Psi({\bf p'}) = {\cal E}\Psi({\bf
p}).
\end{equation}
Here ${\cal U}({\bf p})=\int d{\bf r}e^{-i{\bf pr}}U(r)=2\pi
\int_0^\infty dr rU(r) J_0(pr)$ is the Fourier-transform of the
potential ($J_0(z)$ is the Bessel function). Because of the axial
symmetry of the problem  it is possible to  separate the
cylindrical harmonics of the spinor wave function and search for
the solution in the form
 \begin{eqnarray} \label{spinor}
\mbox{$\Psi^{(m)}$}({\bf p}) =  \left(
\begin{array}{c} \psi_1^{(m)}(p)e^{im\varphi} \\
 \psi_2^{(m)}(p)e^{i(m+1)\varphi} \end{array} \right),
\end{eqnarray}
($\varphi$ is the azimuthal angle of the vector ${\bf p}$). Using
the summation theorem \cite{beit} $$J_0(|{\bf p-p'}|r) =
\sum_{k=-\infty}^\infty J_0(pr)J_0(p'r)\cos{(k\theta)}$$ ($\theta$
is the angle between the vectors ${\bf p}$ and ${\bf p'}$) one can
rewrite Eq.(\ref{Schreq_p}) for each $m$-th harmonic:
\begin{eqnarray} \label{psiC}
[\frac{p^2}{2m_e} -{\cal E}]\psi_{1,2}^{(m)}(p) \pm i\alpha
p\psi_{2,1}^{(m)}(p)+\int_0^\infty dr' r'U(r')J_{m+(1\mp
1)/2}(pr')C_{1,2}^{(m)}(r')=0.
\end{eqnarray}
Here the functions $C_{1,2}^{(m)}$ have been introduced:
\begin{equation}\label{C}
C_1^{(m)}(r)=\int_0^\infty
dppJ_m(pr)\psi_1^{(m)}(p),~~C_2^{(m)}(r)=\int_0^\infty
dppJ_{m+1}(pr)\psi_2^{(m)}(p).
\end{equation}
 Resolving
Eq.(\ref{psiC}) with regard to $\psi_{1,2}$ we find:
\begin{eqnarray} \label{psiC1}
  \left(
\begin{array}{c} \psi_1^{(m)}(p) \\
 \psi_2^{(m)}(p) \end{array} \right) = -\frac{1}{\Delta(p;{\cal E})}\int_0^\infty dr r U(r)
 \left( \begin{array}{c} [p^2/2m_e -{\cal E}]J_m(pr)C_1^{(m)}(r)
 -i\alpha pJ_{m+1}(pr)C_2^{(m)}(r)\\ i\alpha
 pJ_{m}(pr)C_1^{(m)}(r) + [p^2/2m_e -{\cal
 {\cal E}}]J_{m+1}(pr)C_2^{(m)}(r)
  \end{array}\right),
\end{eqnarray}
where $\Delta(p;{\cal E})=(p^2/2m_e -{\cal E})^2-4\alpha^2p^2$.
Zeros of $\Delta(p;{\cal E})$ as  functions of ${\cal E}$  give
two branches of the dispersion relation  for  free electrons:
${\cal E}_\pm(p) =p^2/2m_e \pm \alpha p$.

Finally, from Eq.(\ref{psiC1}) using the definitions (\ref{C}) we
arrive at the equations for $C_{1,2}^{(m)}$
\begin{equation}\label{eqC}
C_i^{(m)}(r)=\int_0^\infty dr'r' U(r')  A_{ij}^{(m)}(r,r')
C_j^{(m)}(r') ~~~~~~~~~~~~~~~~~~ (i,j = 1,2).
\end{equation}
Here  the  matrix $\hat{A}_{ij}^{(m)}$  has been introduced:
\begin{eqnarray} \label{A}
A_{ii}^{(m)}(r,r') = -\int_0^\infty \frac{dp p}{\Delta(p;{\cal
E})}(\frac{p^2}{2m_e}-{\cal E})J_{m+i-1}(pr)J_{m+i-1}(pr'),
\\ \label{A1}  A_{12}^{(m)}(r,r') = -i\alpha\int_0^\infty \frac{dp
p^2}{\Delta(p;{\cal E})}  J_{m}(pr)J_{m+1}(pr'),~~~~~~~
A_{21}^{(m)}(r,r')=(A_{12}^{(m)}(r',r))^*.
\end{eqnarray}

The function $\Delta(p;{\cal E})$ can be presented in the form \
$\Delta(p;E)=[(p-p_0)^2/2m_e-E]\cdot[(p+p_0)^2/2m_e-E]$, where
$p_0=m_e|\alpha|$ is the radius of the loop of extrema, \ $E$ is
the energy counted from the bottom of continuum, $E={\cal E}
+m_e\alpha^2/2$. Now we search for levels of negative energy
satisfying the condition $|E|\ll m_e\alpha^2$ and simultaneously
we assume $2m_eU_0R^2/\hbar^2 \equiv \xi \ll 1$  \ ($U_0, R$ are
the characteristic depth and radius of the well). Then integrals
in Eqs.(\ref{A},\ref{A1}) can be calculated in the "pole"
approximation: we put $p=p_0 $ everywhere in the integrand except
the first factor in $\Delta(p;E)$. As a result we have
($\hbar=1$):
\begin{eqnarray} \label{eqC1}
C_1^{(m)}(r)= -\frac{\pi p_0\sqrt{m}}{\sqrt{2|E|}}\int_0^\infty
dr' r'U(r') [J_m(p_0r)]J_m(p_0r')C_1^{(m)}(r') \nonumber \\
-i~ \mbox{sign}(\alpha)J_m(p_0r)]J_{m+1}(p_0r')C_2^{(m)}(r')],
\nonumber
\\ C_2^{(m)}(r)= -\frac{\pi p_0\sqrt{m_e}}{\sqrt{2|E|}}\int_0^\infty dr'
r'U(r')
[i~\mbox{sign}(\alpha)J_{m+1}(p_0r)]J_m(p_0r')C_1^{(m)}(r')
\nonumber \\ +J_{m+1}(p_0r)]J_{m+1}(p_0r')C_2^{(m)}(r')]
\end{eqnarray}
Thus, we obtained the system of linear integral equations with
degenerate kernels which can be easily solved. This system can be
reduced to a pair of linear algebraic equations for the quantities
$t_{m}\equiv \int_0^\infty dr rU(r)J_m(p_0r)$ and $t_{m+1}$
(defined similarly):
\begin{eqnarray} \label{eqt}
t_m= \frac{\chi_m}{\sqrt{2|E|}}[t_m-i~\mbox{sign}(\alpha)t_{m+1}],
\nonumber
\\ t_{m+1}=
\frac{\chi_{m+1}}{\sqrt{2|E|}}[t_{m+1}+i~\mbox{sign}(\alpha)t_{m}].
\end{eqnarray}
Here $\chi_m = -\pi p_0\sqrt{m_e}\int_0^\infty dr
rU(r)J_m^2(p_0r)$. From Eq.(\ref{eqt}) one immediately gets
\begin{eqnarray}\label{E}
 E_m=-(\chi_m+\chi_{m+1})^2/2 =-\frac{\pi^2 p_0^2 m_e}{2}\left(\int_0^\infty
 dr~rU(r)[J_m^2(p_0r)+J_{m+1}^2(p_0r)]\right )^2 \nonumber
 \\ =-\frac{\pi^2 p_0^2 m_e}{2}\left(\int_0^\infty
 dr~rU(r)[J_{j-1/2}^2(p_0r)+J_{j+1/2}^2(p_0r)]\right )^2,
\end{eqnarray}
where $j=m+1/2=\pm 1/2,\pm 3/2 ...$ is the $z$-projection of the
total moment. As it is seen from Eq.(\ref{E}) all levels are
two-fold degenerate: $E_m$ is even function of $j$.
   If now SO interaction is small ($p_0R\ll 1$) we can get the asymptotic
behavior of the binding energy by expanding the Bessel functions
in Eq.(\ref{E}). For a rectangular well $U(r)=-U_0\theta(R-r)$  we
have $|E_m|\propto
\alpha^{4|j|}/(2^{4|j|}((|j|-1/2)!)^4(2|j|+1)^2)$. For an
exponential well $U(r)=-U_0\exp(-r/R)$ one can find: $|E_m|\propto
\alpha^{4|j|}((2|j|)!)^2/(2^{4|j|}((|j|-1/2)!)^4)$.

Thus, we see that in an arbitrary axially symmetric short-range
(the integral in Eq.(\ref{E})  converges) potential well there
exists at least one bound state for each cylindrical harmonic with
the energy level below the bottom of continuum ($-m_e\alpha^2/2$).
The energy of this state $E$ counted from $-m_e\alpha^2/2$ in the
regime $|E| \ll m_e\alpha^2$ is proportional to $U_0^2$, where
$U_0$ is the characteristic depth of the well. Such a dependence
is typical for a shallow level in a symmetric 1D potential well.
One-dimensional character of the motion results from the so called
"loop of extrema" (see \cite{rash}). In a small vicinity of the
bottom of continuum the dispersion law of 2D electrons has a form
${\cal E}(p)=-m_e\alpha^2/2+(p-p_0)^2/2m_e$ and corresponds to a
1D particle at least in the sense of the density of states: one
may formally consider the problem as the motion of a particle with
anisotropic effective mass;  in the ${\bf p}$-space the radial
component of the mass equals $m_e$, while its azimuthal component
is infinitely large (the dispersion law is independent of the
angle in ${\bf p}$-plane).\cite{raikh}

We realize that our conclusion looks paradoxically: for a
sufficiently large value of $m$ the centrifugal barrier (CB) can
make the effective potential energy  $U(r)+U_{CB}$ positive all
over the space. How can a bound state with {\it negative } energy
be formed in such a situation? Our arguments are as follows: for a
particle with dispersion relation $(p-p_0)^2/2m_e$ there exists no
CB; the azimuthal effective mass tends to infinity and CB
vanishes.

To check our results we have numerically analyzed the square well
potential $U(r)=-U_0\theta(R-r)$ where $\theta$ is the Heaviside
function. We seek for a solution of Schr\"{o}dinger equation in
the form
\begin{eqnarray}
\label{spinor1} \Psi(r,\varphi) = \left(
\begin{array}{c} \psi_1(r)~e^{im\varphi} \\
 \psi_2(r)~e^{i(m+1)\phi} \end{array} \right),
        \end{eqnarray}
        where now $\varphi$ is the azimuthal angle of the vector
        ${\bf r}$.
 Spinor components $\psi_{1,2}(r)$ are given by linear
combinations of the Bessel functions
$J_m(\tilde{\kappa}_{\pm}r),~J_{m+1}(\tilde{\kappa}_{\pm}r)$
 for $r < R$ or $K_m(\kappa_{\pm}r),~
K_{m+1}(\kappa_{\pm}r)$       for $r > R$, where
\begin{eqnarray}\label{kappa}
\tilde{\kappa}_{\pm} = \sqrt{2m_e(E
+U_0)}\pm m\alpha; \nonumber \\
\kappa_{\pm} = \sqrt{2m_e|E|}\pm im\alpha.
\end{eqnarray}
Expressions (\ref{kappa}) are valid when the condition $E < 0$ is
satisfied. Now we have to meet the matching conditions for the
wave function and its derivative at $r=R$. After rather cumbersome
algebra we arrive at the determinants, zeros of which give the
required spectrum of localized states. The energy levels have been
estimated numerically for s- and p-states (m=0, 1). The results
totaly coincide with the ones given above for  $|E| \ll m_e\alpha
^2$. Fig.1 demonstrates this for s-state. The exited $p$-state at
zero SO interaction appears when $U_0$ exceeds a certain critical
value $U_0^{(c)}$, namely, when $\xi >\xi_c=x_1^2$, where $x_1$ is
the first root of the Bessel function $J_0(x)$. Taking into
account  SO interaction results in splitting of the $p$-level and
lowering the critical value $U_0^{(c)}$ for the upper of
spin-split sublevels. The lower sublevel exists at any value of
the parameter $\xi$ (see Fig.2).

 \begin{figure}
\includegraphics[width=14cm]{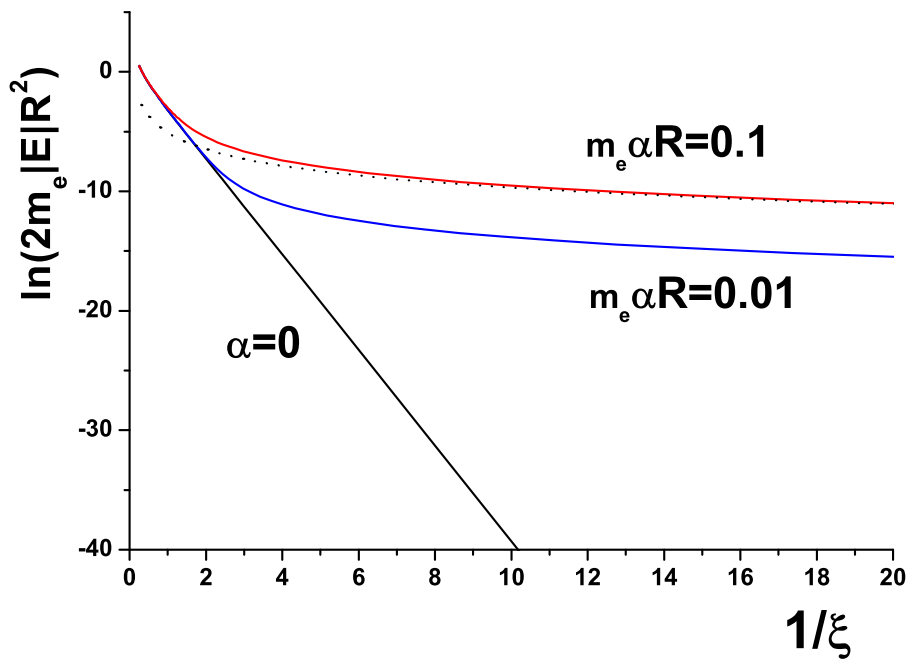}
\caption{\label{fig1} The behavior of the s-level versus the well
depth.  The curves demonstrate the transition between 2D and 1D
regimes. At $\alpha=0$ we have an exponentially shallow level (2D
result), while for finite $\alpha$ at small enough $\xi\equiv
2m_eU_0R^2$ the binding energy parabolically depends on $U_0$  (1D
regime). The dotted line represents the results of our "pole"
approximation (Eq.(\ref{E})) }
\end{figure}

\begin{figure}
\includegraphics[width=14cm]{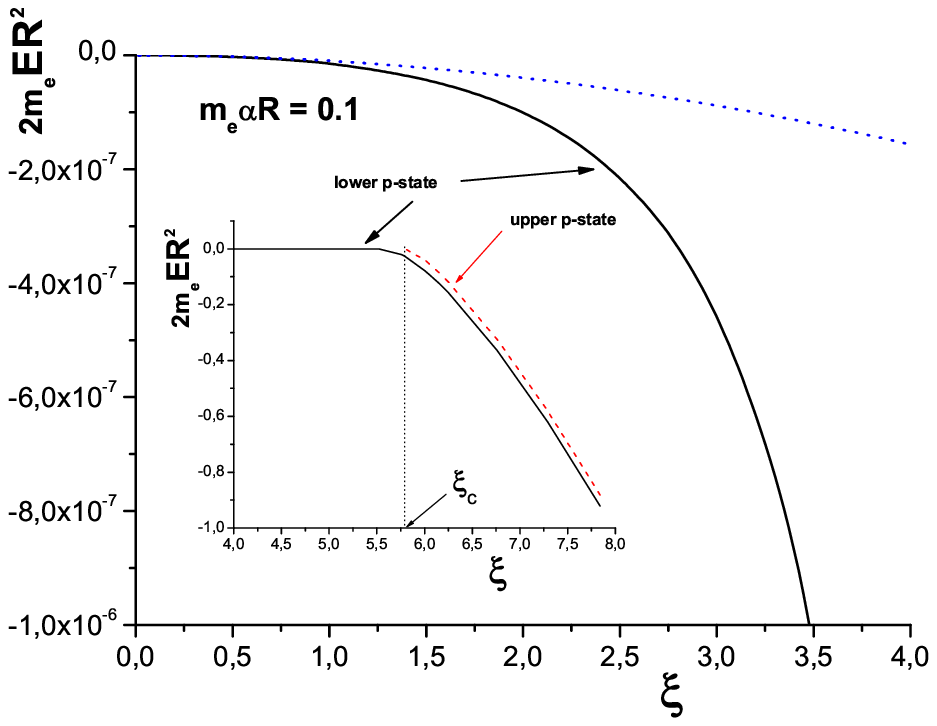}
\caption{\label{fig2} $p$-states. Comparison of the exact solution
for the square well with Eq.(\ref{E}) (dotted line). Inset: spin
split states of the $p$-level: the upper curve terminates  at
$U_0=U_0^{(c)}$ - the level merges  with continuum.}
\end{figure}

 Our last remark relates to  the Coulomb interaction with a charged
impurity. If one tries to apply the general relation (\ref{E}) to
the Coulomb potential $-e^2/r$  the integral logarithmically
diverges at the upper limit: $J_m^2(z\to\infty) \sim (2/\pi
z)\cos^2{(z-\pi m/4-\pi/4)}$ does not depend on $m$ after
averaging over oscillations and can be replaced by $1/\pi z$.
Eq.(\ref{E}) gives for the energy the value independent of $m$:
\begin{equation}\label{EC}
                          E=-2m_ee^4\ln^2{(p_0L)},
\end{equation}
where $L$ is some cut-off length. Its exact value depends on the
concrete situations: it may be the screening radius or the
thickness of 2DEG.
        We see that the energy spectrum does not depend on $m$ (as it must be
for the Coulomb field) and exactly coincides with that of a "1D
hydrogen atom": the ground state binding energy equals $2Ry$
(rather than $Ry/2$ as in 3D case) multiplied $ \ln^2(\Lambda)$,
where $\Lambda$ is the cut-off parameter (see, for example, the
problem of the hydrogen atom in an extremely high magnetic field
\cite{landau}). This result also supports our interpretation: in
the region $|E|\ll m_e\alpha^2$ the particle becomes effectively
one-dimensional.

 It is interesting from the general physics point of view to find a
similar situation for  3D case. E.G.Batyev has kindly reminded us
that the roton spectrum of liquid He-4 also contains a part of
dispersion relation that reads  $\Delta + (p - p_0)^2 /2M $,
possesses not a loop but a surface of extrema, and,
correspondingly, should describe an effectively 1D particle. We
have made the proper calculation, in other words we solved the
Schrodinger equation in the momentum representation for the
Hamiltonian $\Delta + ( p - p_0)^2 /2M  + U(r)$ with $U(r)$ as an
attractive spherically symmetric potential. We used the same
method - expansion of the wave function over the spherical
harmonics and we got the same result: even in 3D a shallow
potential well contains one bound state for each moment $l$ and
this state is $(2l +1)$-fold degenerate:
      \begin{equation}\label{Erot}
E=-2\pi^2p_0^2M\left(\int_0^\infty
dr~rU(r)J_{l+1/2}^2(p_0r)\right)^2.
\end{equation}
For the Coulomb potential the last formula once again leads to
the 1D result given by Eq.(\ref{EC}).

   In conclusion, we have shown that 2D electrons interact with impurities
by a very special way if one takes into account SO coupling: due
to the loop of extrema the system behaves as 1D one for  negative
energies close to the bottom of continuum. This results in the
infinite number of bound states even for a short range potential.

\begin{acknowledgments}
   We thank M.V.Entin for numerous valuable comments and useful discussions.
This work has been  supported by the RFBR grant No 05-02-16939, by
the Council of the President of the Russian Federation for Support
of  Leading Scientific Schools (project no. NSh-593.2003.2),
 and by the Programs of the Russian Academy of
 Sciences.
\end{acknowledgments}

\end{document}